\documentclass{aa}  

\usepackage{graphicx}
\usepackage[parfill]{parskip}
\usepackage{txfonts}
\usepackage[separate-uncertainty = true,multi-part-units=single,range-units=single]{siunitx}

\begin{document}

\title{Tracing magnetic switchbacks to their source: An assessment of solar coronal jets as switchback precursors}

\titlerunning{Tracing magnetic switchbacks to their source}

  \author{N. Bizien\inst{\ref{inst1}}
         \and
         C. Froment \inst{\ref{inst1}} 
         \and
         M. S. Madjarska \inst{\ref{inst2},\ref{inst3}, \ref{inst4}}
         \and
         T. Dudok de Wit \inst{\ref{inst1},\ref{inst5}}
         \and 
         M. Velli\inst{\ref{inst6}}}

  \institute{LPC2E, OSUC, Univ Orleans, CNRS, CNES, F-45071 Orleans, France\\
             \email{nina.bizien@cnrs-orleans.fr}
            \label{inst1} 
   \and
     Max Planck Institute for Solar System Research, Justus-von-Liebig-Weg 3, 37077, G\"ottingen, Germany
      \label{inst2}
      \and
      Korea Astronomy and Space Science Institute, 34055, Daejeon, Republic of Korea
      \label{inst3}
      \and
       Space Research and Technology Institute, Bulgarian Academy of Sciences, Acad. G. Bonchev Str., Bl. 1, 1113, Sofia, Bulgaria
       \label{inst4}
   \and
            International Space Science Institute, 3012, Bern, Switzerland
            \label{inst5}
   \and
     Earth, Planetary \& Space Sciences, University of California, Los Angeles, CA, USA 
       \label{inst6}}

   \date{Received xxx; accepted xxx}

  \abstract
   {The origin of large-amplitude magnetic field deflections in the solar wind, known as magnetic switchbacks, is still under debate. These structures, which are ubiquitous in the in situ observations made by Parker Solar Probe (PSP), likely have their seed in the lower solar corona, where small-scale energetic events driven by magnetic reconnection could provide conditions ripe for either direct or indirect generation.}
   {We investigated potential links between in situ measurements of switchbacks and eruptions originating from the clusters of small-scale solar coronal loops known as coronal bright points to establish whether these eruptions act as precursors to switchbacks.}
   {We traced solar wind switchbacks from PSP back to their source regions using the ballistic back-mapping and potential field source surface methods, and analyzed the influence of the source surface height and solar wind propagation velocity on magnetic connectivity. Using extreme ultraviolet images, we combined automated and visual approaches to identify small-scale eruptions (e.g., jets) in the source regions. The jet occurrence rate was then compared with the rate of switchbacks captured by PSP.}
   {We find that the source region connected to the spacecraft varies significantly depending on the source surface height, which exceeds the expected dependence on the solar cycle and cannot be detected via polarity checks. For two corotation periods that are straightforwardly connected, we find a matching level of activity (jets and switchbacks), which is characterized by the hourly rate of events and depends on the size of the region connected to PSP. 
   However, no correlation is found between the two time series of hourly event rates. Modeling constraints and the event selection may be the main limitations in the investigation of a possible correlation.
   Evolutionary phenomena occurring during the solar wind propagation may also influence our results. These results do not allow us to conclude that the jets are the main switchback precursors, nor do they rule out this hypothesis. They may also indicate that a wider range of dynamical phenomena are the precursors of switchbacks.}
   {}

\maketitle
%
\section{Introduction}
\label{sec_introduction}

Switchbacks are magnetic field deflections with associated velocity jets and have attracted considerable attention because of their potential role in mediating the heating and acceleration of the solar wind. Their origin, however, remains elusive. They were first detected by \textit{Ulysses} \citep{balogh_heliospheric_1999}, but Parker Solar Probe \citep[PSP;][]{fox_solar_2016} measurements have revealed their ubiquity \citep[e.g.,][]{bale_highly_2019,kasper_alfvenic_2019}. Their origin is still debated: switchbacks could be produced in situ in the solar wind via turbulent processes or by the expansion of the solar wind \citep{Landietal:2006,ruffolo_shear-driven_2020,squire_-situ_2020, toth_theory_2023}. Other interpretations suggest a low coronal origin through, for instance, interchange magnetic reconnection \citep[e.g.,][]{fisk_global_2020, drake_switchbacks_2021,schwadron_switchbacks_2021}. There is, however a growing consensus on the key role played by small-scale energetic events in creating the conditions necessary for the generation of switchbacks through magnetic reconnection.

Switchbacks are ubiquitous in the solar wind; therefore, we investigated events that can act as potential precursors regardless of the phase of the solar cycle and the presence of active regions or coronal holes (CHs). The spatial and temporal ubiquity of reconnection-driven small-scale eruptions such as coronal jets \citep[e.g.,][]{shibata_observations_1992, raouafi_solar_2016}, most of which originate from clusters of small-scale loops known as coronal bright points \citep[CBPs;][]{madjarska_coronal_2019}, makes them good candidates for triggering switchbacks. Jets from CBPs manifest as collimated plasma flows along open or closed magnetic field lines with speeds often as high as a few hundred kilometers per second. A subpopulation of jets can escape from the solar corona, following open magnetic field lines \citep[e.g.,][]{1998ApJ...508..899W,nistico_characteristics_2009,paraschiv_study_2010}. The common hypothesis goes beyond the simple propagation of kinked field lines, created during interchange reconnection, and involves the magnetic untwisting of waves that can survive until they reach PSP \citep{touresse_propagation_2024}.

The direct comparison of recurrent eruptive phenomena and switchbacks is challenging. While several studies have focused either on the in situ properties of switchbacks \citep[e.g.,][]{bale_interchange_2023} or on remote sensing of their possible solar sources \citep[e.g.,][]{huang_statistical_2023}, here we combine the two approaches in an effort to investigate a causal connection between the two phenomena. 
Periodicity comparisons have been performed to probe the connectivity between solar transient events and magnetic structure \citep{kumar_new_2023, hou_connecting_2024}. To avoid potential pitfalls in the interpretation, we further developed this hypothesis by applying more stringent conditions for event selection, such as unambiguous connectivity and the use of synchronous observations. Establishing a causal link between eruptive solar phenomena in the low corona and the switchbacks (observed in situ) triggered by them presents several challenges. 
The first is to establish the magnetic connectivity; such a determination will be highly dependent on the input data and is sometimes based on nonsynchronous observations \citep{de_pablos_searching_2022}, increasing the risk of error. Uncertainties also arise from disentangling spatial and temporal phenomena during observation intervals, such as longitudinal sweeps.

Here we explore the hypothesis of jets as a switchback source in the lower corona by combining in situ and remote-sensing observations, using a methodology that includes an analysis of the influence of source surface heights and the propagation velocity to connect the two measurements. We include and discuss all the possible known sources of uncertainties from the current instrumental and modeling constraints.
\section{Data and methods}\label{sec_data}
Our analysis focuses on corotation intervals of PSP with the Sun, which increases the probability of measuring the same source over a long period and gives greater confidence in the connectivity. One significant advantage is that it allows for co-temporal analysis as we attempt to compare transient events.

Switchbacks are identified from in situ magnetic field measurements made by the MAG magnetometer on board PSP \citep{bale_fields_2016}. We considered sharp deflections from the Parker spiral as in \citet[see Fig~\ref{fig:ex_catalog} for an illustration of the identification methodology; see also Dudok de Wit, in prep and \citealt{dudok_de_wit_switchbacks_2020}]{bizien_are_2023}. We excluded events that lasted less than \SI{3}{s} and small events with a deflection of less than \SI{36}{\degree}, as they are difficult to distinguish from random fluctuations.

Our connectivity investigation is based on a standard two-step methodology involving back-mapping and a magnetic connectivity model. First, we performed ballistic back-mapping to determine the radial evolution of the solar wind measured by PSP back to a source surface \citep{nolte_large-scale_1973}. We determined how long it takes for perturbations to propagate from the Sun and evolve into switchbacks detected at PSP. For this we relied on the reconstructed proton velocity data product from the Solar Probe Analyzer-Ion instrument of the Solar Wind Electrons Alphas and Protons suite (SWEAP/SPAN-I) on board PSP \citep{livi_solar_2022}.

The velocity at PSP is computed using a median over $\pm1$~h interval with a constant velocity profile (from the source surface to the spacecraft). We tested three different hypotheses to compare the jet and switchback rates: a constant velocity profile, a more realistic profile with acceleration, and a profile with acceleration that accounts for the local Alfvén velocity. Using the hybrid profile of \citet{koukras_estimating_2022} and the isopoly model \citep{dakeyo_statistical_2022}, the propagation time of an acceleration profile is found to be equivalent to the one calculated with a constant profile with a correction of $-80~\mathrm{km~s}^{-1}$. The propagation of Alfvén waves produced by jets can be estimated using the model presented in \citet{collier_reconnection_2001}. The propagation time computed for an acceleration profile that also includes the Alfvén velocity model is found to be equivalent to adding $80~\mathrm{km~s}^{-1}$ to the constant velocity.

We also computed the full connectivity for velocities measured at PSP, adding an interval of velocities $\pm80~\mathrm{km~s}^{-1}$ with increments of $10~\mathrm{km~s}^{-1}$ \citep{panasenco_exploring_2020,bale_solar_2021}. The increments account for the correction of the uncertainties in the velocity measurements. The correction also compensates for the use of a constant velocity profile over acceleration profiles, which is estimated to cause a difference of \SIrange{4}{14}{\degree} on the identification sources in the Carrington maps \citep{dakeyo_radial_2024,koukras_estimating_2022}.

Next, we traced the identified point on the source surface to the Sun using a potential-field source-surface (PFSS) extrapolation model \citep[][]{altschuler_magnetic_1969, stansby_pfsspy_2020}. We used photospheric magnetic field data from the Air Force Data Assimilative Photospheric Flux Transport \citep[ADAPT;][]{arge_air_2010} modeled magnetogram based on Global Oscillation Network Group (GONG) magnetograms \citep[][]{harvey_global_1996}.

The largest influence on the identified source region comes from the choice of the source surface height  \citep{koukras_estimating_2022,ervin_characteristics_2024}, which is usually set at $2.5 R_{\sun}$. A variability due to the selection of the source surface is expected, as it also depends on the solar cycle phase \citep{levine_open_1977,schulz_coronal_1978,levine_simulation_1982}. A selection of a unique height for the entire encounter is generally made based on the best match of polarities at the source surface and in situ \citep{badman_constraining_2022}. A local estimation can also be performed, varying at each polarity change \citep{panasenco_exploring_2020}. However, polarity crossing alone is insufficient for the specific connectivity to PSP, given the short temporal and spatial scales of the events we are investigating. Thus, we explored surface heights between $1.5R_{\sun}$ and $3.5 R_{\sun}$, using magnetograms taken every two hours.

To investigate the solar sources of the switchbacks, we used extreme ultraviolet images at \SI{193}{\angstrom} taken by the Atmospheric Imaging Assembly \citep[AIA;][]{lemen_atmospheric_2012} of Solar Dynamics Observatory \citep[SDO;][]{pesnell_solar_2012}, hereafter referred to as AIA~193. We analyzed eruptive CBPs on a data cube with de-rotated images -- temporally binned -- at a cadence of 36~s. We performed a visual identification to detect the transient eruptive CBPs in the source region. To ease the detection of smaller-scale events and those lacking background contrast, we used an automatic transient brightening identification procedure that identifies intensity increases above a local background threshold \citep[for details see][]{subramanian_coronal_2010}. 

%
\section{Results and discussion}
\label{sec_results} 
We considered the first seventeen solar encounters of PSP, focusing on three periods out of 33 when PSP was close to perihelion and mostly corotating with the Sun. Two encounters (E10 and E15) provide straightforward connectivity between PSP and the Sun. 
However, this is an exception rather than the rule: in most other cases, we find multiple possible source regions depending on the height of the source surface, highlighting the inherent limitations of the PFSS modeling. This exceeds the predicted variability as the connection can move from an equatorial CH at a lower source surface to a polar CH at a higher source surface, maintaining the same polarity. An example of a lack of uniqueness is shown in Appendix~\ref{appendix_cases} for E05.

\begin{figure}
    \centering
    \includegraphics[width=\hsize]{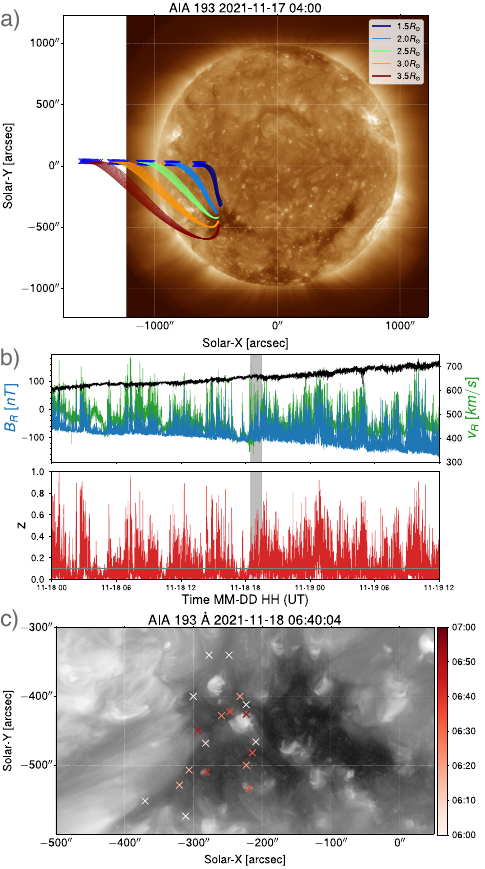}
    \caption{E10 connectivities. (a): Context AIA~193 image taken at the time of the computed extrapolation model. Colored magnetic field lines extracted from the PFSS model, computed at different source surface heights, are overplotted. The blue cross markers are the projection of PSP on the different source surfaces. The multiple blue field lines correspond to different velocities. All field lines are rooted in the same equatorial CH.  (b): In situ PSP data for E10 for a constant Carrington longitude. The $z$ parameter corresponds to the normalized deflection from the Parker spiral. The gray horizontal line shows the lower threshold for the deflection angle. The gray vertical span corresponds to the time that matches the identification presented in panel (c), assuming propagation with constant velocity. (c): Zoomed-in view of the CH connected to PSP during the entire corotating interval, with the jets detected during a 1-hour interval.} 
    \label{fig:case1}
    \end{figure}

\subsection{Connectivity during E10}

The outbound corotating interval E10 occurred from 2021 November 18 at 00:00 to November 19 at 12:00 UT \citep{shi_patches_2022}. The back-mapping time estimation gives a corresponding solar interval from 2021 November 17 at 11:00 to November 19 at 00:00 UT.

    \begin{figure}
    \centering
    \includegraphics[width=\hsize]{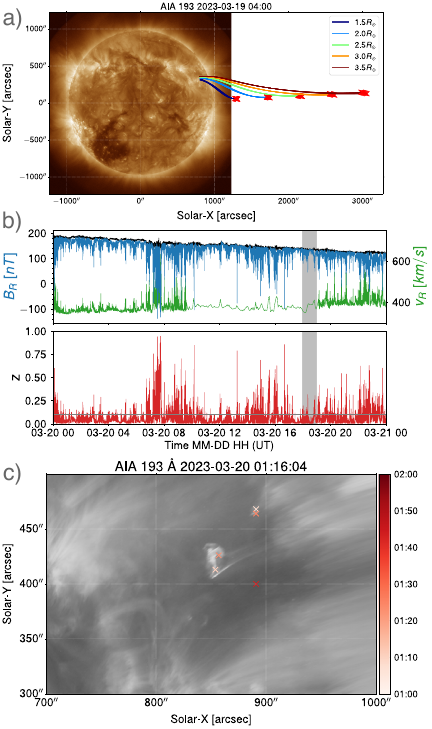}
    \caption{Same as Fig.~\ref{fig:case1} but for E15. PSP is connected at the beginning of the corotating interval to a small, near-limb CH.} 
    \label{fig:case2}
    \end{figure}

PSP is connected with a CH, shown in Fig.~\ref{fig:case1}a with an AIA 193 context image taken at the time of the PFSS. The predicted negative polarity of the photosphere matches the polarity observed at PSP (radial magnetic field in Fig.~\ref{fig:case1}b). The different colored lines show the multiple velocity calculations, with the color indicating different source surface heights. The crosses are the projection of PSP onto the source surfaces. The connectivity agrees with the results of \citet{badman_prediction_2023}, who studied the entire E10 and in particular the interval we are considering.
The connectivity of all field lines gives the same source region. The computations for the different velocities do not influence the area connected to the spacecraft. However, the different source surface heights result in different subregions in the CH with varying latitudes, as shown in Fig.~\ref{fig:case1}c and consistent with previous studies \citep{dakeyo_radial_2024, koukras_estimating_2022}. 

Figure~\ref{fig:case1}b shows the in situ solar wind magnetic field and proton velocity measurements at PSP during the corotating interval. It corresponds to a fast wind interval, which is consistent with the expected properties of solar wind escaping from a CH. We find 508 switchbacks that comply with our selection criteria (see Sect.~\ref{sec_data} and Fig.~\ref{fig:ex_catalog}). 

The jets we identify have a minimum base size of 3\arcsec\ (2.17 Mm) and a minimum duration of 3~min. Figure~\ref{fig:case1}c presents the eruptions counted during a 1-hour interval on November 18, from 06:00 to 07:00 UT, and they are marked with the starting time color-coded. We considered the same subregion of the CH for all 1-hour intervals.

    \begin{figure*}
    \centering
    \includegraphics[width=\hsize]{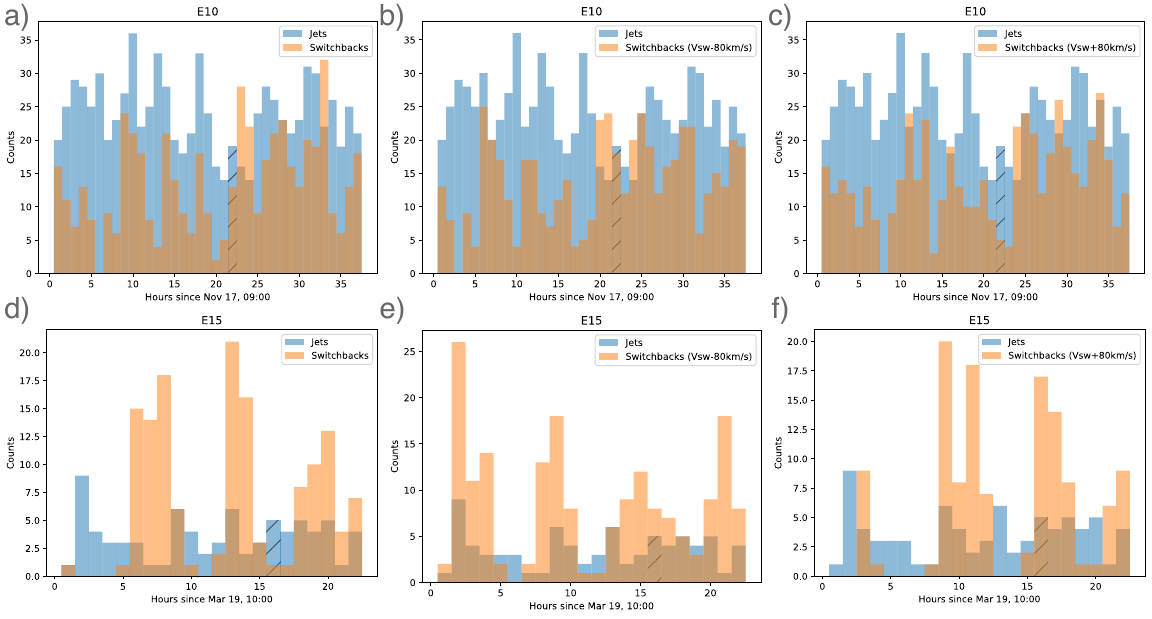}
    \caption{Histogram of jet and switchback counts. The hourly counts of jets are shown in blue and the corresponding counts of switchbacks in orange. The reference time corresponds to the start of the jet-counting interval. (a): E10 counts with constant velocity. 
    (b): E10 counts with acceleration correction. (c): E10 counts with acceleration correction and Alfv\'en velocity. Each switchback bin, which is equivalent to the 1~h interval of solar eruptions, corresponds to counts over 50~min. (d), (e), and( f): Same as panels (a), (b), and (c) but for E15. The hashed bins correspond to the counts in Figs.~\ref{fig:case1}c and \ref{fig:case2}c, respectively. Each switchback bin, which is equivalent to the 1~h interval of CBP eruptions, corresponds to counts over 1h10m.}
    \label{fig:correlations}
    \end{figure*}
    
\subsection{Connectivity during E15}
The outbound corotating interval E15 occurred from 2023 March 20 at 00:00 to March 21 at 12:00 UT. The corresponding solar interval is from 2023 March 19 to March 20. The connectivity of the spacecraft with the Sun is within a small CH close to the solar limb, near NOAA~13252, as noted two days earlier (Fig.~\ref{fig:case2}a). The CH polarity is positive, matching the polarity at PSP. Neither the input back-mapped velocity nor the source surface height changes the connectivity.

Figure~\ref{fig:case2}b shows the in situ measurements at PSP during the corotating interval. The activity level is lower than E10, as we detect only 142 switchbacks during this interval, with fewer large deflections. The solar wind velocity is lower, about $400~\mathrm{km~s}^{-1}$. We considered the jets escaping from the CBPs in the region connected to PSP. Figure~\ref{fig:case2}c shows the counts of jets during a 1-hour interval starting on March 20, at 01:00 UT.

\subsection{Correlation analysis of jets and switchbacks}\label{subsec:res_counts}

Interestingly, there is a matching level of activity for the two cases, with comparable high rates of jets (649/day) and switchbacks (346/day) for the large CH of E10 and comparable low rates of jets (79/day) and switchbacks (142/day) activity for the smaller CH of E15. Thus, the switchback occurrence varies with the source region properties \citep{fargette_characteristic_2021}.  

We then tested the three hypotheses regarding the propagation velocity described in Sect.~\ref{sec_data}, as it is a critical parameter influencing the back-mapping time and the comparison of the two datasets. We used the same models as for the connectivity computation, which resulted in the same velocity corrections.

We show in Fig.~\ref{fig:correlations}a for E10 the hourly rate of eruptive events compared to that of switchbacks, for the period from November 17 at 09:00 to November 18 at 22:00 UT. 
To begin with the first hypothesis, we used a constant velocity profile as a first approximation to estimate the travel time of the eruptive events, excluding wind acceleration.
The value of the Pearson's correlation coefficient between the rates of jets and switchbacks is $r=0.05$. This is smaller than the threshold value $r^*= \pm 0.32$ obtained for a level of $\alpha=0.05$, so we conclude that there is no significant linear correlation.

The second hypothesis involves a lower velocity to account for the longer propagation time associated with the acceleration profile. This would assume that the jets have a similar acceleration profile as the solar wind. The ``$-80~\mathrm{km~s}^{-1}$'' acceleration-corrected histogram plot, presented in Fig.~\ref{fig:correlations}b, has a statistically significant anticorrelation of $r=-0.42$.

Finally, the third hypothesis is to account for the Alfv\'en waves produced during the jets \citep{lionello_contribution_2016}, directly linked to the Alfv\'enic properties of the switchbacks \citep{kasper_alfvenic_2019}. Thus, one could include the Alfv\'en velocity in the computation of the propagation time. We present  in Fig.~\ref{fig:correlations}c the results with acceleration, including the local Alfv\'en velocity. The correlation ($r=0.03$) is statistically not significant.

Figure~\ref{fig:correlations}d compares the hourly rates of jets and switchbacks for E15, using the same setup as in E10. Here again, there is no apparent  correlation, and indeed the value of the correlation ($r=-0.10$) is smaller than the threshold value $r^*=\pm 0.42$ obtained for a level of $\alpha=0.05$. When including the correction of the acceleration profile, as in Fig.~\ref{fig:correlations}e, the correlation becomes statistically significant with a value of $r=0.44$. For the test case, which accounts for the acceleration profile and the local Alfv\'en velocity in Fig.~\ref{fig:correlations}f, the correlation ($r=0.03$) is statistically not significant.

These results reveal how much the level of correlation depends on the assumed velocity profile, which defines the shift between the two records. We tested various different velocities in the uncertainty range of $V \in [-80, +80] ~\mathrm{km~s}^{-1}$. The values of the correlation coefficient are generally not significant but vary considerably. This is partly a consequence of events appearing in clusters lasting about five hours (visual evidence of clustering). Shifting the time series by 2.5 hours then reverses the value of $r$. The occasional presence of statistically significant anticorrelations, such as in E10, which are physically unrealistic, cautions us against taking statistically barely significant values of $r$ as evidence of a causal relationship between the jets and the switchbacks.

\section{Sources of uncertainties}\label{sec_discussion}
We used a methodology to compare the input and output of a complex system involving several physical mechanisms, each adding another layer of uncertainty. We accounted for most possible sources of uncertainties, but some remain unresolved and need further study. We discuss their potential impacts below. These uncertainties tend to be multiplicative, making it very challenging to establish a causal or even statistical connection between the input and the output.

\subsection{Magnetic connectivity uncertainties}

Estimating the connectivity is critical in establishing the causal relation between dynamical phenomena and switchbacks. However, it relies on simple extrapolation, assuming a potential field, the simplest case of a force free-corona. This is significantly important for small-scale, dynamically evolving structures. Moreover, the input magnetogram resolution, about \SI{1}{\degree}, is low compared to the size of the events (the footprint size is 30\arcsec\ across on average). Another issue is that the connectivity is based on global synoptic maps, not only synchronous data, which affects accuracy when active regions appear on the Sun's far side \citep{perri_impact_2024}. 
The PFSS modeling is very restrictive in the rate of events we can confidently study in the estimated source region. The source is statistical, which does not account for reconnection in the solar wind that can bring structures from other sources \citep{viall_nine_2020}. Magnetohydrodynamic modeling is not expected to improve this analysis due to the difficulty of tracking individual events. 

\subsection{Complexity of the jetting activity}

A visual inspection of Fig.~\ref{fig:correlations} suggests the presence of matching clustering in jet and switchback rates, indicating that some switchbacks may indeed have been produced during the jets' generation. The clustering may be associated with the switchback patches \citep[e.g.,][]{bale_interchange_2023}. The lack of a clear correlation and the jet rate being lower than the switchback rate suggests that some switchbacks may not be initiated in the low corona but at higher coronal heights or in situ in the solar wind. Additional mechanisms may also be contributing to the formation of switchbacks. \citet{hou_origin_2024} show a direct link between jets and switchbacks using a similar methodology. Their study, however, is limited to the assumed uncertainties in the propagation behavior. Therefore,  a direct link between a group of switchbacks and one jet remains to be investigated more rigorously when new suitable data become available.

The jetting activity adds further complexity to the interpretation. We may be underestimating the rate of eruptions in the source region due to limitations in the visual identification in extreme ultraviolet images, such as dark and/or stealth jets \citep[e.g.,][]{young_dark_2015}, or due to multiple eruptions happening at the location of one jet \citep{madjarska_dynamics_2011, chen_recurrent_2015}. The apparent open field, which may be considered as plumes, is located outside the CH, while the PFSS gives the source in the CH, so we cannot consider the influence of the jetlets \citep{kumar_new_2023}. We might also be overestimating the rate of jets that could have a causal link observed in the in situ record since it is impossible to state that all jets escape the corona. The size of the source region investigated also plays a role in the selection, due to the latitudinal uncertainties of the source surface height, and would directly affect the rate of detected jets (e.g., a larger CH where we observe more CBPs but with a greater spatial variability, as in Case 1).
The different sizes and energetics of the jets and possibly different acceleration profiles of the produced Alfv\'enic fluctuations may also strongly influence the comparative analysis. Uncertainties in the evolutionary phenomena occurring during solar wind propagation may also play a significant role.

The temporal clustering leads directly to significant temporal variability in the rate of jets from the same region; this is also evident in the occurrence of switchbacks, which is possibly associated with switchback patches \citep{fargette_characteristic_2021}. This highlights the often overlooked temporal variability of in situ switchback properties \citep{shi_patches_2022}. This variability persists for events with smaller deflections. The event selection (jets and switchbacks) likely influences the observed variability.

Unlike \citet{hou_connecting_2024}, we did not perform a periodicity analysis since we cannot rule out the possibility that the event-counting and the combination of spatial and temporal effects influence some of the identified clusters. The spatial effects would arise because the spacecraft would not maintain a consistent connection to a single flux tube over the entire interval. The physical cause of a recurring pattern in the jets from a large region warrants further investigation, as it is not straightforward.

\section{Summary and conclusions} \label{sec_conclusion}

We analyzed the connectivity between the Sun and PSP and used a jet identification approach to explore the causal connection between switchbacks and small-scale dynamic phenomena, that is, jets associated with clusters of small-scale loops (CBPs). The initial analysis of all available corotation periods (33 over 17 encounters) revealed a greater sensitivity of the connectivity to the source surface height than expected from the literature \citep[e.g.,][]{badman_constraining_2022, schulz_coronal_1978}: extreme variations (over 60 Carrington degrees in latitude) of the solar source connected to the spacecraft due to the source surface height selection that exceed the variation due to the solar cycle. 

In most cases, multiple regions of open magnetic flux, as given by the PFSS model, could be connected to the spacecraft. In addition, the standard connectivity method has significant limitations that affected our analysis of small-scale transient events associated with the local opening of the magnetic field. 

Therefore, we only considered those few cases for which unambiguous connectivity could be found. We observed a matching level of activity (jets and switchbacks) in some cases; but, the level of activity varied depending on the size of the CH connected to PSP. However, we did not find a clear temporal correlation between the rate of jets from the source region and the rate of switchbacks at PSP. The event selection, which depends on the subregion considered, the instrument's capability to detect some jets (such as dark jets), and the threshold on switchback deflection angle, may have significantly affected these results. Lastly, our interpretation might be influenced by our limited understanding of the evolutionary processes occurring during solar wind propagation: the results do not allow us to conclude that the jets are the main precursors of switchback, nor do they rule out this hypothesis.

Taken together, we identify many cumulative uncertainties, and while this does not imply that we cannot test the connection, it does caution us. Therefore, establishing a causal one-to-one connection between solar jets and switchbacks at PSP has not yet been achieved. 

The physical phenomenon considered (the bulk plasma flow or the Alfv\'en waves) influences the results by means of the propagation velocity. The time shift for the correspondence of the time series has a direct impact on the correlation. However, this is highly dependent on the temporal modulation of the switchbacks observed during corotating intervals with the same source region. It shows that the variability of switchbacks has both spatial and temporal components, as argued by \cite{shi_patches_2022}. The similar variability in the jets raises new questions since no temporal modulation of jets is expected.

Future work should focus on refining the jet identification to avoid binning limitations. The connectivity issues could be overcome by using more complex and combined extrapolation models that account for the small scales and transient nature of the events. This may allow one to investigate intervals in which few or no switchbacks are observed, which could provide more insight into the link between jets and switchbacks.

\begin{acknowledgements}
We thank the anonymous referee for their thorough review of the manuscript.
The FIELDS experiment was developed and is operated under NASA contract NNN06AA01C. N.B., T.D., and C.F. acknowledge the financial support of the CNES in the frame of a Parker Solar Probe grant. C.F. acknowledges funding from the CEFIPRA Research Project No. 6904-2. M.M. acknowledges support from the University of Orl\'eans, as a visiting Professor, DFG grant WI 3211/8-2, project number 452856778, and the support of the Brain Pool program funded by the Ministry of Science and ICT through the National Research Foundation of Korea (RS-2024-00408396). The authors thank Srividya Subramanian for providing the latest version of the brightening identification code. N.B. thanks Jaye Verniero and Kristoff Paulson for their advice about SPAN-I and SPC data. N.B. thanks Fr\'ed\'eric Auch\`ere for the Python module for the correction of the differential rotation of EUV images.  Parker Solar Probe was designed and built and is now operated by the Johns Hopkins Applied Physics Laboratory as part of NASA’s Living with a Star (LWS) program (contract NNN06AA01C). Support from the LWS management and technical team has played a critical role in the success of the Parker Solar Probe mission.
The PSP data used in this study are available at the NASA Space Physics Data Facility (SPDF),  \href{https://spdf.gsfc.nasa.gov}{https://spdf.gsfc.nasa.gov}.
SDO is a mission for NASA's LWS program. AIA is an instrument on board the Solar Dynamics Observatory. All SDO data used in this work are available from the Joint Science Operations Center \href{http://jsoc.stanford.edu}{http://jsoc.stanford.edu} without restriction.
This analysis relies on the \textit{pfsspy} implementation of PFSS \citep{stansby_pfsspy_2020}.
\end{acknowledgements}

%
\bibliographystyle{aa} 
\bibliography{biblio1.bib} 
%

\begin{appendix}

\section{Jet count}

This section presents the jets identified over a one-hour interval for both periods considered in the paper. Table \ref{table:counts_e10} indicates the jets identified during the interval 2021 November 18 from 06:00~UT to 07:00 of E10. It corresponds to the jets marked in Fig.~\ref{fig:case1}c). Table \ref{table:counts_e15} indicates the jets during the interval 2023-03-20 from 01:00 to 02:00 of E15. It corresponds to the jets marked in Fig.~\ref{fig:case2}c).
\begin{table}[h]
\caption{Jets identified during the interval 2021 November 18 from 06:00 to 07:00 of E10.} 
\label{table:counts_e10} 
\centering 
\begin{tabular}{| c | c | c |} 
\hline 
Jets & Location [X,Y] in arcseconds & Start (UTC) \\ 
\hline 
1& [-300,-400]&06:00\\
2 &[-277,-340]& 06:00\\
3& [-223,-412]&06:00\\
4&[-311,-574]&06:01\\
5&[-370,-552]&06:02\\
6&[-248,-340]&06:04\\
7&[-209,-466]&06:06\\
8&[-282,-468]&06:07\\
9&[-306,-507]&06:14\\
10&[-320,-529]&06:15\\
11&[-232,-400]&06:18\\
12&[-259,-428]&06:23\\
13&[-223,-500]&06:23\\
14&[-214,-482]&06:33\\
15&[-247,-422]&06:35\\
16&[-219,-534]&06:36\\
17&[-281,-510]&06:40\\
18&[-293,-449]&06:50\\
19&[-223,-426]&06:54\\
\hline 
\end{tabular}
\tablefoot{The location and the time of the jet's start are indicated. It corresponds to the jets marked in Fig.~\ref{fig:case1}c) for illustration. This is a sample for the sake of conciseness. }
\end{table}

\begin{table}[h]
\caption{Jets identified during the interval 2023 March 20 from 01:00 to 02:00 of E15. } 
\label{table:counts_e15} 
\centering 
\begin{tabular}{| c | c | c |} 
\hline 
Jets & Location [X,Y] in arcseconds & Start (UTC) \\ 
\hline 
1 & [891,468] &  01:08 \\ 
2 & [891,464] &  01:25\\
3 & [857,426] &  01:23\\
4 & [854,413] &  01:11\\
5 & [891,400] &  01:43\\

\hline 
\end{tabular}
\tablefoot{The location and the time of the start of the jet are indicated. It corresponds to the jets marked in Fig.~\ref{fig:case2}d). }
\end{table}

\section{Identification of switchbacks}

\begin{figure*}[h!]
    \centering
    \includegraphics[width=\linewidth]{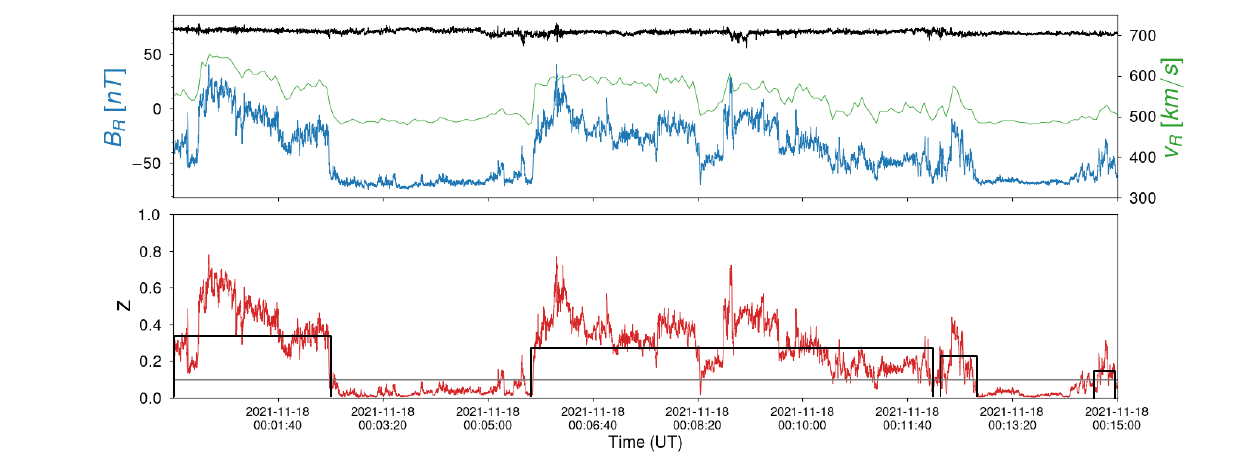}
    \caption{Example of a switchback interval. Their identification is indicated with the black rectangular shapes. Top: FIELDS/MAG magnetic field data and the proton velocity from SWEAP/SPAN-I, both in the R direction. Bottom: Normalized deflection, $z$. The overlapping rectangular shapes represent the structures included in this analysis that are sudden and/or sharp deflections.}
    \label{fig:ex_catalog}
\end{figure*}

This section presents an example of a sub-selection of switchbacks included in this analysis. We identified switchbacks with a minimal deflection of \SI{36}{\degree} and a minimal duration of 3s. Figure~\ref{fig:ex_catalog} shows four switchbacks in the magnetic field measured by PSP, with the rectangular shape showing the model fit.

\section{Case 3 analysis}\label{appendix_cases}
While the two cases presented in Sect.~\ref{sec_results} were straightforward in interpreting the connectivity given by the PFSS, it is not the same for most of the PSP corotating intervals. The outbound corotating interval of E05 is more complex and actually more representative of the PSP encounters. 
It occurred from 2020 June 10 to June 11 at 06:00. The emission time starts on 2020 June 08 at 18:00 and ends on 2020 June 09 at 22:00.

\begin{figure}[h!]
    \centering
    \includegraphics[width=\hsize]{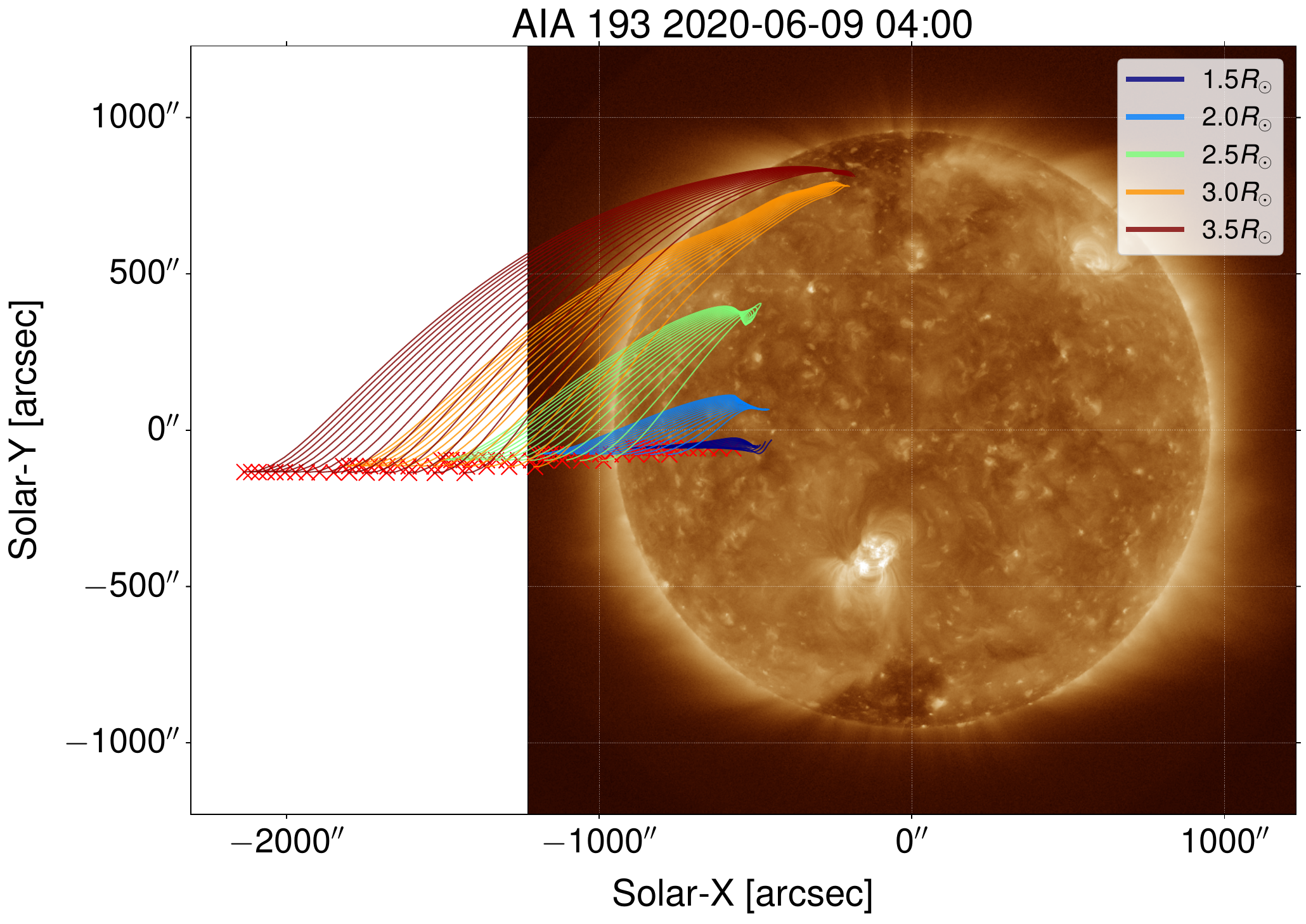}
    \caption{Connectivity at the beginning of the E05 corotating interval for Case 3. It shows the inherent limitations of the conclusions on the connectivity, which depend on the source surface choice for some cases.}
    \label{fig:e05_connect_start}
    \end{figure}

Figure~\ref{fig:e05_connect_start} shows one possible connectivity output of E05 corotation, which depends strongly on the source surface height selection. Not all the footpoints are located in the same CH. Lower heights of the source surface will give a footpoint located in a small CH, a higher source surface give a footpoint in the quiet Sun, while an even higher source surface shows connectivity in the northern polar CH. 

From the available PSP encounters, after removing the events when the PSP is not connected to the visible part of the disk, only the two cases presented above show an unambiguous connectivity region. The remaining encounters are similar to Case 3, with jumping connectivity and multiple possible source regions depending on the source surface height. This highlights the inherent limitations of the PFSS.

\end{appendix}

\end{document}